\documentclass[12pt,a4paper]{article}
\pdfoutput=1

\usepackage[utf8]{inputenc}
\usepackage[T1]{fontenc}
\usepackage[margin=2.0cm]{geometry}
\usepackage[pdfstartview={FitH},bookmarks=false,linktoc=page,colorlinks=true,linkcolor=blue,citecolor=blue,urlcolor=blue]{hyperref}
\usepackage[numbered]{bookmark}
\usepackage{mathtools}
\usepackage{amssymb}
\usepackage{graphicx}
\usepackage[font=small,labelfont=bf]{caption}
\usepackage{authblk}
\usepackage{cite}
\usepackage{dsfont}
\usepackage{xcolor}
\usepackage[titles]{tocloft}
\usepackage{float}
\usepackage{subcaption}
\usepackage{color}
\usepackage{tikz}
\usepackage{ifthen}
\usepackage[warn]{textcomp}
\numberwithin{equation}{section}
\allowdisplaybreaks
\usepackage{multirow}

\begin{document}

\title{\vspace{2cm}\textbf{Information Geometry on the Space of Equilibrium States of Black Holes in Higher Derivative Theories}\vspace{1cm}}

\author[a,b]{Tsvetan Vetsov}

\affil[a]{\textit{Department of Physics, Sofia University,}\authorcr\textit{5 J. Bourchier Blvd., 1164 Sofia, Bulgaria}

\vspace{-10pt}\texttt{}\vspace{0.0cm}}

\affil[b]{\textit{The Bogoliubov Laboratory of Theoretical Physics, JINR,}\authorcr\textit{141980 Dubna,
Moscow region, Russia}

\vspace{10pt}\texttt{vetsov@phys.uni-sofia.bg}\vspace{0.1cm}}
\date{}
\maketitle

\begin{abstract}
We study the information-geometric properties of the Deser-Sarioglu-Tekin black hole, which is a higher derivative gravity solution with contributions from a non-polynomial term of the Weyl tensor to the Einstein–Hilbert Lagrangian. Our investigation is focused on deriving the relevant information metrics and their scalar curvatures on the space of equilibrium states. The analysis is conducted within the framework of thermodynamic information geometry and shows highly non-trivial statistical behavior. Furthermore, the quasilocal formalism, developed by Brown and York, was successfully implemented in order to derive the mass of the Deser-Sarioglu-Tekin black hole.
\end{abstract}

\vspace{0.5cm}
\vspace{0.5cm}
\thispagestyle{empty}

\noindent\rule{\linewidth}{0.75pt}
\vspace{-0.8cm}\tableofcontents
\noindent\rule{\linewidth}{0.75pt}

\section{Introduction}\label{sec: Introduction}

In recent years the puzzling existence of dark matter and dark energy, which cannot be explained either by Einstein's General theory of relativity (GR) or the Standard model of elementary particles, suggests that alternative models have to be kept in mind. 
\\
\indent From gravitational perspective one can consider modified theories of gravity and in particular higher derivative theories (HDTs), which include contributions from polynomial or non-polynomial functions of the scalar curvature. The most prominent of them are the so called $f(R)$ theories \cite{STAROBINSKY198099, DeFelice:2010aj, Cembranos:2008gj}. The $f(R)$ gravity is a whole family of models with a number of predictions, which differ from those of GR. Therefore, there is a great deal of interest in understanding the possible phases and stability of such higher derivative theories and, thereof, their admissible 
black hole solutions \cite{Cognola:2005de, Cognola:2006eg, Oliva:2010eb, Oliva:2010zd, Cai:2010zh, Berezhiani:2008nr}.
\\
\indent However, a consistent description of black holes necessarily invokes the full theory of quantum gravity. Unfortunately, at present day, our understanding of such theory is incomplete at best. This prompts one to resort to alternative approaches, which promise to uncover many important aspects of quantum gravity and black holes. One such example is called  information geometry \cite{amari2007methods, Amari:2016:IGA:3019383, amari2012differential, ay2017information}.
\\
\indent The framework of information geometry is an essential tool for understanding how classical and quantum information can be encoded onto the degrees of freedom of any physical system. Since geometry studies mutual relations between elements, such as distance and curvature, it provides us with a set of powerful analytic tools to study previously inaccessible features of the systems under consideration. It has emerged from studies of invariant geometric structures arising in statistical inference, where one defines a Riemannian metric, known as Fisher information metric \cite{amari2007methods}, together with dually coupled affine connections on the manifold of probability distributions. Information geometry already has important applications in many branches of modern physics, showing intriguing results. Some of them, relevant to our study, include condensed matter systems \cite{Wh0, Wh1, PhysRevA.20.1608, Janyszek:1989zz, Johnston:2003ed, Dolan2655, Dolan:2002wm, Janke02, Janke:2002ne, Quevedo:2015xdx, Quevedo:2008hh, Quevedo:2013bze, Quevedo:2010tz, vetsov2018}, black holes \cite{Aman:2003ug, Shen:2005nu, Janke10, Ferrara:1997tw, Cai:1998ep, Aman:2005xk, Aman:2007pp, Aman:2007ae, Suresh:2016cqp, Suresh:2016onn, Quevedo:2016swn, Channuie:2018mkt, Quevedo:2016cge, Larranaga:2010kj, Sarkar:2006tg, Astefanesei:2010bm, Mansoori:2016jer, Mansoori:2014oia, Mansoori:2013pna, e20030192, Miao:2018fke, Ruppeiner:2018pgn} and string theory \cite{Ferrara:1997tw, Larranaga:2010kj, Dimov:2017ryz, Dimov:2016vvl}. Further applications can also be found in \cite{Amari:2016:IGA:3019383, ay2017information}.
\\
\indent When dealing with systems such as black holes, which seem to possess enormous amount of entropy \cite{PhysRevD.7.2333, bardeen1973, PhysRevD.9.3292, hawking1972}, one can consider their space of equilibrium states, equipped with a suitable Riemannian metric such as the Ruppeiner information metric \cite{RevModPhys.67.605}. The latter is a thermodynamic limit of the above-mentioned Fisher information metric. Although G. Ruppeiner developed his geometric approach within fluctuation theory, when utilized for black holes, it seems to capture many features of their phase structure, resulting from the dynamics of the underlying microstates. In this case one implements the entropy as a thermodynamic potential to define a Hessian metric structure on the state space statistical manifold with respect to the  extensive parameters of the system. 
\\
\indent Moreover, one can utilize the internal energy (the ADM mass in the case of black holes) as an alternative thermodynamic potential, which lies at the heart of Weinhold's metric approach \cite{Wh0} to equilibrium thermodynamic states. The resulting Weinhold information metric is conformally related to Ruppeiner metric, with the temperature $T$ as the conformal factor. Unfortunately, the resulting statistical geometries coming from both approaches do not often agree with each other. The reasons for this behavior are still unclear, although several attempts to resolve this issue have already been suggested \cite{Sarkar:2006tg, Mirza:2007ev, Quevedo:2017tgyy, Quevedo:2017tgz, Mansoori:2016jer}.
\\
\indent In the current paper we are going to study the equilibrium thermodynamic state space of the Deser-Sarioglu-Tekin (DST) black hole \cite{Deser:2007za} within the framework of thermodynamic information geometry. The DST black hole solution is a static, spherically symmetric black hole solution in higher derivative theory of gravity with contributions from a non-polynomial term of the Weyl tensor to Einstein–Hilbert Lagrangian. 
\\
\indent The text is organized as follows. In Section \ref{sec:2} we shortly discuss the basic concepts of geometrothermodynamics and related approaches. In Section \ref{sec:3} we calculate the mass of the DST solution via the quasilocal formalism developed by Brown and York in \cite{Brown:1992br}. In Section \ref{sec:4} we calculate the standard thermodynamic quantities such as the entropy and the Hawking temperature of the DST black hole solution and we show that the first law of thermodynamics is satisfied. In Sections \ref{sec:5} and \ref{sec:6} we study the Hessian information metrics and several Legendre invariant approaches, respectively. We show that the Hessian approaches of Ruppeiner and Weinhold fail to produce viable state space metrics, while the Legendre invariant metrics successfully manage to incorporate the Davies phase transition points. Finally, in Section \ref{sec:conclusion}, we make a short summary of our results.

\section{Information geometry on the space of equilibrium thermodynamic states}\label{sec:2}
Due to the pioneering work of Bekenstein \cite{PhysRevD.7.2333} and Hawking \cite{hawking1972} we know that any black hole represents a thermal system with well-defined temperature and entropy. Taking into account that black holes may also possess charge $Q$ and angular momentum $J$, one can formulate the analogue to the first law of thermodynamics for black holes as
\begin{equation}\label{eqBHFirstLaw}
dM = T\,dS + {\Phi _Q}\,dQ + \Omega \,dJ\,.
\end{equation}
Here $\Phi _Q$ is the electric potential and $\Omega$ is the angular velocity of the event horizon. Equation (\ref{eqBHFirstLaw}) expresses the conserved ADM mass $M$ as a function of entropy and other extensive parameters, describing the macrostates of the black hole. One can equivalently solve Eq. (\ref{eqBHFirstLaw}) with respect to the entropy $S$. 
\\
\indent In the framework of geometric thermodynamics all extensive parameters of the given black hole background can be used in the construction of its equilibrium thermodynamic parameter space. The latter can be equipped with a Riemannian metric in several ways. In particular, one can introduce Hessian metrics, whose components are calculated as the Hessian of a given thermodynamic potential. For example, depending on which potential we have chosen  for the description of the thermodynamic states in equilibrium, we can write the two most popular thermodynamic metrics, namely the Weinhold information metric \cite{Wh0},
\begin{equation}
ds_W^2 = {\partial _a}{\partial _b}M\,d{X^a}\,d{X^b}\,,
\end{equation}
defined as the Hessian of the ADM mass $M$, or the Ruppeiner information metric \cite{PhysRevA.20.1608},
\begin{equation}\label{eqRinfometric}
ds_R^2 =  - {\partial _a}{\partial _b}S\,d{Y^a}\,d{Y^b}\,,
\end{equation}
defined as the Hessian of the entropy $S$. Here $X_a [Y_b], a,\,b=1,\dots,\,n,$ collectively denote all of the system's extensive variables except
for $M [S]$.
One can show that both metrics are conformally related to each other via the temperature:
\begin{equation}
ds_W^2 = T\,ds_R^2\,.
\end{equation}
The importance of using Hessian metrics on the equilibrium manifold is best understood when one considers small fluctuations of the thermodynamic
potential. The latter is extremal at each equilibrium point, but the second moment
of the fluctuation turns out to be directly related to the components of the corresponding Hessian metric. From statistical point of view one can define  Hessian metrics on a statistical manifold spanned by any type or number of extensive (or intensive) parameters. In this case the first law of thermodynamics has to be properly generalized in order to include the chemical potentials of all relevant fluctuating parameters. This is due to the fact that the Hessian metrics are not Legendre invariant, thus they do not necessarily preserve the geometric properties of the system when a different thermodynamic potential is chosen. However, for Legendre invariant metrics, the first law of thermodynamics follows naturally.
\\
\indent In order to make things Legendre invariant, one can start from the $(2n + 1)$-dimensional thermodynamic phase space $\mathcal{F}$, spanned by the thermodynamic potential $\Phi$, the set of extensive variables $E^a$, and the set of intensive variables $I^a$, $a = 1, \dots, n$. Now, consider a symmetric bilinear form $\mathcal{G}=\mathcal{G}(Z^A)$ defining a non-degenerate metric on $\mathcal{F}$ with $Z^A=(\Phi,\,E^a,\,I^a)$, and the Gibbs 1-form $\Theta  = d\Phi  - {\delta _{ab}}\,{I^a}\,d{E^b}$, where $\delta_{ab}$ is the identity matrix. If the condition $\Theta\wedge(d\Theta)^n\neq 0$ is satisfied, then the triple ($\mathcal{F},\,\mathcal{G},\,\Theta$) defines a contact Riemannian manifold. The Gibbs 1-form is invariant with respect to Legendre transformations by construction, while the
metric $\mathcal{G}$ is Legendre invariant only if its functional dependence on $Z^A$ does not change under a Legendre transformation. Legendre invariance guarantees that the geometric properties of $\mathcal{G}$ do not depend on the choice of thermodynamic potential.
\\
\indent On the other hand, one is interested in constructing a viable Riemannian metric $g$ on the $n$-dimensional  subspace of equilibrium thermodynamic states $\mathcal{E}\subset \mathcal{F}$. The space $\mathcal{E}$ is defined by the smooth mapping $\phi:\mathcal{E}\to\mathcal{F}$ or $E^a\to(\Phi(E^a),\,E^a,\,I^a)$, and the condition $\phi^*(\Theta)=0$. The last restriction leads explicitly to the generalization of the first law of thermodynamics (\ref{eqBHFirstLaw})
\begin{equation}\label{eqGeneralizedFirstLawofTD}
d\Phi  = {\delta _{ab}}\,{I^a}\,d{E^b}\,,
\end{equation}
and the condition for thermodynamic equilibrium,
\begin{equation}
\frac{{\partial \Phi }}{{\partial {E^a}}} = {\delta _{ab}}\,{I^b}\,.
\end{equation}
The natural choice for $g$ is the pull-back of the phase space metric $\mathcal{G}$ onto $\mathcal{E}$, $g=\phi^*(\mathcal{G})$. Here, the pull-back also imposes the Legendre invariance of $\mathcal{G}$ onto $g$. However, there are plenty of Legendre invariant metrics on $\mathcal{F}$ to choose from. In Ref.  \cite{Quevedo:2017tgz} it was found that the general metric for
the equilibrium state space can be written in the form
\begin{equation}
{g^{I,\,II}} = {\beta _\Phi }\,\Phi ({E^c})\,\chi _a^{\,\,b}\frac{{{\partial ^2}\Phi }}{{\partial {E^b}\,\partial {E^c}}}\,d{E^a}\,d{E^c}\,,
\end{equation}
where $\chi^{\,\,b}_a=\chi_{af}\,\delta^{fb}$ is a constant diagonal matrix and $\beta_\Phi\in\mathbb{R}$ is the degree of generalized homogeneity, $\Phi ({\lambda ^{{\beta _1}}}\,{E^1}, \ldots ,{\lambda ^{{\beta _N}}}\,{E^N})=$ ${\lambda ^{{\beta _\Phi }}}\,\Phi ({E^1}, \ldots ,{E^N}),\,{\beta _a} \in \mathbb{R}$. In this case the Euler's identity for homogeneous functions can be generalized to the form
\begin{equation}\label{eqEulerIdentity}
{\beta _{ab}}\,{E^a}\,\frac{{\partial \Phi }}{{\partial {E^b}}} = {\beta _\Phi }\,\Phi \,,
\end{equation}
where ${\beta _{ab}} = {\mathop{\rm diag}\nolimits} ({\beta _1},\,{\beta _2}, \ldots ,{\beta _N})$. In the case $\beta_{ab}=\delta_{ab}$ one returns to the standard Euler's identity. If we choose to work with $\beta_{ab}=\delta_{ab}$, for complicated systems this may lead to non-trivial conformal factor, which is no longer proportional to the potential $\Phi$. On the other hand, if we set $\chi_{ab}=\delta_{ab}$, the resulting
metric $g^I$ can be used to investigate systems with at least one first-order phase transition. Alternatively, the choice $\chi_{ab}=\eta_{ab}={\rm{diag}}(-1,1,\dots,1)$ leads to a metric $g^{II}$, which applies to systems with second-order phase transitions.
\\
\indent Once the information metric for a given statistical system is constructed, one can proceed with calculating its algebraic invariants, i.e. the information curvatures such as the Ricci scalar, the Kretschmann invariant, etc. All curvature related quantities are relevant for extracting information about the phase structure of the system. As suggested by G. Ruppeiner in Ref. \cite{RevModPhys.67.605}, the Ricci information curvature $R_I$ is related to the correlation volume of the system. This association follows from the idea that it will be less probable to fluctuate from one equilibrium thermodynamic state to the other, if the distance between the points on the statistical manifold, which correspond to these states, increases. Furthermore, the sign of $R_I$ can be linked to the nature of the inter-particle interactions in composite thermodynamic systems \cite{2010AmJPh..78.1170R}. Specifically, if $R_I=0$, the interactions are
absent, and we end up with a free theory (uncorrelated bits of information). The latter situation corresponds to flat information geometry. For positive curvature, $R_I>0$, the interactions are repulsive, therefore we have an elliptic information geometry, while for negative curvature, $R_I<0$, the interactions are of attractive nature and an information geometry of hyperbolic type is realized. 
\\
\indent Finally, the scalar curvature of the parameter manifold can also be used to measure the stability of the physical system under consideration. In particular, the information curvature approaches infinity in the vicinity of critical points, where phase transition occurs \cite{Janyszek:1989zz}. Moreover, the curvature of the information metric tends to diverge not only at the critical points of phase transitions, but on whole regions of points on the statistical space, called spinodal curves. The latter can be used to discern physical from non-physical situations. 
\\
\indent Furthermore, notice that in the case of Hessian metrics, in order to ensure global thermodynamic stability of a given
macro configuration of the black hole, one requires that all principal minors of the metric tensor be strictly positive definite, due to the probabilistic interpretation involved \cite{PhysRevA.20.1608}. In any other cases (Quevedo, HPEM, etc) the physical interpretation of the metric components is unclear and one can only impose the convexity condition on the thermodynamic potential, $\partial_a\partial_b\Phi\ge 0$, which is the second law of thermodynamics. Nevertheless, imposing positiveness of the black hole's heat capacity is mandatory in any case in order to ensure local thermodynamic stability.

\section{The DST black hole}\label{sec:3}
One starts with the following action (in units $\kappa = 1$)
\begin{equation}\label{eqTheDSTaction}
A = \frac{1}{{2}}\,\int_{\cal M} {{d^4}x\,\sqrt { - g} \,\left( {R + {\beta _n}\,|{\mathop{\rm Tr}\nolimits} {(C^n)}{|^{1/n}}} \right)}\,,
\end{equation}
where $C$ is the Weyl tensor and $\beta_n$ are some real constant coefficients. The spherically symmetric Deser-Sarioglu-Tekin solution \cite{Deser:2007za, Bellini:2010ar},
\begin{equation}\label{eqDSTmetric}
d{s^2} =  - {k^2}\,{r^{\frac{{2\,(1 - p(\sigma ))}}{{p(\sigma )}}}}\,\left( {p(\sigma ) - \frac{c}{{{r^{1/p(\sigma )}}}}} \right)\,d{t^2} + \frac{{d{r^2}}}{{p(\sigma ) - \frac{c}{{{r^{1/p(\sigma )}}}}}} + {r^2}\,(d{\theta ^2} + {\sin ^2}\theta \,d{\phi ^2})\,,
\end{equation}
follows from (\ref{eqTheDSTaction}) by setting $n=2$ and $\sigma=\beta_2/\sqrt{3}$. Here, the integration constant $k$ can be eliminated by a proper rescaling of the time coordinate $t$. For convenience we have defined the function $p(\sigma)$ as
\begin{equation}
p(\sigma ) = \frac{{1 - \sigma }}{{1 - 4\,\sigma }}\,.
\end{equation}
To preserve the signature of the metric, we have to exclude the interval $1/4<\sigma<1$. There is only one horizon of the black hole, which is at the positive root of $g^{rr}=0$:
\begin{equation}
{r_h} = {\left( {c\,\left( {\frac{3}{{\sigma  - 1}} + 4} \right)} \right)^{\frac{{\sigma  - 1}}{{4\,\sigma  - 1}}}}= {\left( {\frac{c}{p}} \right)^p}\,.
\end{equation}
We also note that in general the metric (\ref{eqDSTmetric}) is not asymptotically flat, unless we consider the case $\sigma=0$, for which the charge $c>0$ can be interpreted as the ADM mass of a Schwarzschild black hole ($c=2\,M$). Using the quasilocal formalism \cite{Brown:1992br, Yazadjiev:2005du}, we can support the claim that $M=c/2$ is the mass of the DST black hole for any $\sigma<1/4$ and $\sigma>1$. To show this, one has to bring the DST metric (\ref{eqDSTmetric}) in the form
\begin{equation}\label{eqDSTMetricInYCoord}
d{s^2} =  - \lambda (y)\,d{t^2} + \frac{{d{y^2}}}{{\lambda (y)}} + {R^2}(y)\,d\Omega _2^2\,.
\end{equation}
The following change of variables
\begin{equation}
r = {\left( {\frac{{4\,\sigma  - 1}}{{\sigma  - 1}}\,y} \right)^{\frac{{\sigma  - 1}}{{4\,\sigma  - 1}}}}\,
\end{equation}
is suitable for this task, thus
\begin{equation}\label{eqTheLambdaFunction}
\lambda (y) = (y - c)\,{\left[ {\left( {4 + \frac{3}{{\sigma  - 1}}} \right)y} \right]^{\frac{{2\,\sigma  + 1}}{{4\,\sigma  - 1}}}}\,
\end{equation}
and 
\begin{equation}
{R^2}(y) = {\left( {\frac{{4\,\sigma  - 1}}{{\sigma  - 1}}\,y} \right)^{2\,\frac{{\sigma  - 1}}{{4\sigma  - 1}}}}\,.
\end{equation}
We can now calculate the quasilocal mass, 
\begin{equation}\label{eqQuasiLocalMassGeneral}
{\cal M}(y) = \frac{1}{2}\,\frac{{d{R^2}(y)}}{{dy}}\,{\lambda ^{1/2}}(y)\,\left( {\lambda _0^{1/2}(y) - {\lambda ^{1/2}}(y)} \right)\,,
\end{equation}
derived in \cite{Yazadjiev:2005du}, where $\lambda_0(y)$ is an arbitrary non-negative function which determines the zero of the energy for a
background spacetime. Because there is no cosmological horizon present, the large $y$ limit of (\ref{eqQuasiLocalMassGeneral}) determines the
mass of the black hole. The explicit result for the DST solution is given by
\begin{equation}
{\cal M}(y) = c - y + \sqrt {{\lambda _0}(y)} \,\sqrt {y - c} \,{\left( {\left( {\frac{3}{{\sigma  - 1}} + 4} \right)\,y} \right)^{\frac{{2\,\sigma  + 1}}{{2(1 - 4)\,\sigma }}}}\,.
\end{equation}
The arbitrary function $\lambda_0(y)$ can be fixed as the first term in the large $y$ asymptotic expansion of $\lambda(y)$. The expression is given by
\begin{equation}\label{eqTheLambdaFunction}
\lambda_0 (y) = y\,{\left[ {\left( {4 + \frac{3}{{\sigma  - 1}}} \right)y} \right]^{\frac{{2\,\sigma  + 1}}{{4\,\sigma  - 1}}}}\,
\end{equation}
Now, in the limit $y\to \infty$, one finds the quasilocal mass of the DST black hole,
\begin{equation}\label{eqDSTQLMass}
M = \mathop {\lim }\limits_{y \to \infty } {\cal M}(y) = \frac{c}{2}\,.
\end{equation}
The latter expression relates the unknown integration constant $c$ from Eq. (\ref{eqDSTmetric}) to the quasilocal mass $M$ of the DST black hole. Equation (\ref{eqDSTQLMass}) is valid only when $\frac{{2\,\sigma  + 1}}{{4\,\sigma  - 1}}\ge 0 $, i.e. $\sigma\leq -1/2$ or $\sigma>1/4$.
\\
\indent One can impose further restrictions on the parameter $\sigma$ by calculating the independent curvature invariants, i.e. the Ricci scalar,
\begin{equation}\label{eqDSTRicciPhysical}
R =  - \frac{{6\,\sigma \,\left( {2\,M\,(\sigma  - 1) + 3\,\sigma \,y} \right)}}{{{{(\sigma  - 1)}^{\frac{{2\,\sigma  + 1}}{{4\sigma  - 1}}}}\,{{\left( {(4\,\sigma  - 1)\,y} \right)}^{\frac{{2\,(3\,\sigma  - 1)}}{{4\,\sigma  - 1}}}}}}\,,
\end{equation}
and the Kretschmann invariant,
\begin{align}
K = \frac{{12\,\left( {4\,{M^2}\,\left( {5\,{\sigma ^2} + 1} \right)\,{{(\sigma  - 1)}^2} + 12\,M\,\sigma \,\left( {{\sigma ^3} - 1} \right)\,y + 3\,{\sigma ^2}\,\left( {\sigma \,(7\,\sigma  - 2) + 4} \right)\,{y^2}} \right)}}{{{{(\sigma  - 1)}^{\frac{{2\,(2\,\sigma  + 1)}}{{4\,\sigma  - 1}}}}\,{{\left( {(4\,\sigma  - 1)y} \right)}^{\frac{{6\,(2\,\sigma  - 1)}}{{4\,\sigma  - 1}}}}}}\,.
\end{align}
Both quantities are singular at $\sigma=1$ and $\sigma=1/4$. At $\sigma\to 0$ one recovers the Schwarzschild case.

\section{Thermodynamics of the DST black hole}\label{sec:4}
We proceed with Wald’s proposal \cite{Wald:1993nt} to calculate the entropy, 
\begin{equation}\label{eqWaldEntropyFormula}
S =  - 8\,\pi \,{\oint_{y = {y_h}\hfill\atop
t = const\hfill} {\left( {\frac{{\delta {\cal L}}}{{\delta {R_{ytyt}}}}} \right)} ^{(0)}}\,R(y)\,d{\Omega ^2}\,,
\end{equation}
of the DST black hole with metric in the form (\ref{eqDSTMetricInYCoord}). The variational derivative of the Lagrangian is given by \cite{Bellini:2010ar}:
\begin{align}
\nonumber
\frac{{\delta {\cal L}}}{{\delta {R_{\alpha \beta \gamma \delta }}}}&= \frac{{({g^{\alpha \gamma }}\,{g^{\beta \delta }} - {g^{\alpha \delta }}\,{g^{\beta \gamma }})}}{{32\,\pi }}\,\left( {1 + \frac{{\sqrt 3 \,\sigma \,R}}{{3\,\sqrt {{C^2}} }}} \right)\\
 &+ \frac{{\sqrt 3 \,\sigma }}{{32\,\pi \,\sqrt {{C^2}} }}\left[ {2\,{R^{\alpha \beta \gamma \delta }} - ({g^{\alpha \gamma }}\,{R^{\beta \delta }} + {g^{\beta \delta }}\,{R^{\alpha \gamma }} - {g^{\alpha \delta }}\,{R^{\beta \gamma }} - {g^{\beta \gamma }}\,{R^{\alpha \delta }})} \right]\,.
\end{align}
In Eq. (\ref{eqWaldEntropyFormula}) the superscript (0) indicates that the variational derivative is calculated on the solution. After some lengthy calculations one arrives at the explicit formula for  Wald's entropy of the DST black hole:
\begin{equation}\label{eqDSTEntropy}
S = \pi \,{\left[ {2\,M\,\left( {4 + \frac{3}{{\sigma  - 1}}} \right)} \right]^{\frac{{2\,\left( {\sigma  - 1} \right)}}{{4\sigma  - 1}}}}\,,
\end{equation}
which is positive for $\sigma<1/4$ and $\sigma>1$. The Hawking temperature yields
\begin{equation}
T = \frac{1}{{4\,\pi }}\,\frac{{d\lambda }}{{dy}}({y_h}) = \frac{{{8^{\frac{{1 - 2\,\sigma }}{{4\,\sigma  - 1}}}}}}{\pi }\,{\left[ {M\,\left( {4 + \frac{3}{{\sigma  - 1}}} \right)} \right]^{\frac{{1 + 2\,\sigma }}{{4\,\sigma  - 1}}}}\,,
\end{equation}
where $y_h=2\, M$ is the location of the event horizon given by the zeros of $\lambda(y)=0$. The singularities of the temperature occur at $\sigma=1/4$ and $\sigma=1$. The temperature has one local extremal point at ($M=1/4,\,\sigma=-1/2$), which is a saddle point. At $\sigma\to-1/2$ one has $T\to 1/(4\pi)$ and the DST temperature doesn't depend on the mass $M$. Moreover, the heat capacity,
\begin{equation}\label{eqDSTHeatCapacity}
C = T\,\frac{{\partial S}}{{\partial T}} =  \frac{{\pi \,{8^{\frac{{1 - 2\,\sigma }}{{1 - 4\,\sigma }}}}\,{{(\sigma  - 1)}^{\frac{{2\,\sigma  + 1}}{{4\,\sigma  - 1}}}}{{\left( {M\,(4\,\sigma  - 1)} \right)}^{\frac{{2\,(\sigma  - 1)}}{{4\,\sigma  - 1}}}}}}{{2\,\sigma  + 1}}\,,
\end{equation}
of the DST black hole diverges at $\sigma=1/4$ and $\sigma=-1/2$ and tends to zero at $\sigma=1$. However the points $\sigma=1/4$ and $\sigma=1$ are not physical due to the divergences of the physical curvature (\ref{eqDSTRicciPhysical}), while $\sigma=-1/2$ corresponds to Davies type phase transition. In the limit $\sigma\to 0$ one recovers General relativity, where the heat capacity reduces to the Schwarzschild case, $C=-8 \,\pi\,M^2<0$, which is known to be thermodynamically unstable. Our subsequent considerations will also discard this limit. 
\\
\indent One can also check that the first law of thermodynamics, $dM=T\,dS$, is satisfied.

\section{Hessian thermodynamic geometries on the equilibrium state space of the DST  black hole solution}\label{sec:5}

\subsection{Extended equilibrium state space}
If we consider thermal fluctuations of the parameter $\sigma$, we have to take into account its contribution to the first law of thermodynamics. In Ruppeiner's approach one takes the entropy as a thermodynamic potential, thus
\begin{equation}
dS = \frac{1}{T}\,dM + \Xi \,d\sigma\,,
\end{equation}
which is the generalized first law from Eq. (\ref{eqGeneralizedFirstLawofTD}). Here $\Xi$ plays the role of the chemical potential for $\sigma$ (considered as a new extensive parameter). The explicit form of $\Xi$ is given by
\begin{align}
\Xi  = 3\,\pi \,{8^{\frac{{1 - 2\,\sigma }}{{1 - 4\,\sigma }}}}{(4\,\sigma  - 1)^{\frac{{6\,\sigma }}{{1 - 4\,\sigma }}}}\,{\left( {\frac{M}{{\sigma  - 1}}} \right)^{\frac{{2(\sigma  - 1)}}{{4\sigma  - 1}}}}\,\left[ {\ln \left( {2\,M\,\left( {\frac{3}{{\sigma  - 1}} + 4} \right)} \right) - 1} \right]\,.
\end{align}
The equilibrium thermodynamic state space of the DST solution (\ref{eqDSTmetric}) is now considered as a two-dimensional manifold equipped with a suitable Riemannian metric
\begin{equation}\label{eqInfoMetricGeneral}
ds_I^2 = {g^{(I)}_{ab}}\,d{E^a}\,d{E^b}\,,
\end{equation}
where $E^a$, $a=1,2$, are the extensive parameters such as the mass, the entropy or the parameter $\sigma$. Depending on the chosen thermodynamic potential, one discerns several possibilities for the information thermodynamic metric as discussed in the Introduction section. 
\\
\indent It is also well-known that in 2 dimensions all the relevant information about the phase structure is encoded only in the information metric and its scalar curvature. The latter is proportional to the (only one independent) component of the Riemann curvature tensor,
\begin{equation}\label{eqInfoCurvature2d}
{R_I} = \frac{{2\,{R_{I,1212}}}}{{\det {(g^{(I)}_{ab})}}}\,,
\end{equation}
where $\det ({g^{(I)}_{ab}}) $ is the determinant of the information metric (\ref{eqInfoMetricGeneral}). Once the scalar information curvature is obtained, we can identify its singularities as phase transition points, which should be compared to the resulting divergences of the heat capacity. If a complete match is found one can rely on the considered information metric as suitable for describing the space of equilibrium states for the given black hole solution. 

\subsection{Ruppeiner information metric}

We begin by calculating the Ruppeiner information metric,
\begin{equation}
{g^{(R)}_{ab}} =  - {\partial _a}{\partial _b}{S}(M,\,\sigma),\;\;\;a,b = 1,2\,,
\end{equation}
with components
\begin{align}
g_{MM}^{(R)} &= \frac{{\pi \,{8^{\frac{{2\,\sigma  - 1}}{{4\,\sigma  - 1}}}}\,{{(\sigma  - 1)}^{\frac{{2\,\sigma  + 1}}{{4\,\sigma  - 1}}}}\,(2\,\sigma  + 1)}}{{{{\left( {M\,(4\,\sigma  - 1)} \right)}^{\frac{{6\,\sigma }}{{4\,\sigma  - 1}}}}}}\,,\\\nonumber
\\
g_{M\sigma }^{(R)} &= g_{\sigma M}^{(R)} =  - \frac{{3\,\pi \,{8^{\frac{{2\,\sigma  - 1}}{{4\,\sigma  - 1}}}}\,\left( {2\,(\sigma  - 1)\,\ln \left[ {2\,M\,\frac{{4\,\sigma  - 1}}{{\sigma  - 1}}} \right] + 2\,\sigma  + 1} \right)}}{{{M^{\frac{{2\,\sigma  + 1}}{{4\sigma  - 1}}}}\,{{(\sigma  - 1)}^{\frac{{2\,(\sigma  - 1)}}{{4\,\sigma  - 1}}}}\,{{(4\,\sigma  - 1)}^{\frac{{10\,\sigma  - 1}}{{4\,\sigma  - 1}}}}}}\,,\\\nonumber
\\\nonumber\small
g_{\sigma \sigma }^{(R)} &= \frac{{3\,\pi \,{8^{\frac{{2\,\sigma  - 1}}{{4\,\sigma  - 1}}}}{M^{\frac{{2\,(\sigma  - 1)}}{{4\,\sigma  - 1}}}}}}{{{{(\sigma  - 1)}^{\frac{{2\,(2\,\sigma  - 1)}}{{4\,\sigma  - 1}}}}\,{{(4\,\sigma  - 1)}^{\frac{{2(7\,\sigma  - 1)}}{{4\,\sigma  - 1}}}}}}\,\left\{ {46\,\sigma  - 32\,{\sigma ^2} - 5 + \ln {2^{4\,(8\,{\sigma ^2} - 7\,\sigma  + 1)}} - \ln {8^{\sigma  + 1}}\,\ln 4} \right.\\
 &+ \left. {2\,(\sigma  - 1)\,\left( {2 + 16\,\sigma  - 3\,\ln \left[ {2\,M\,\frac{{4\,\sigma  - 1}}{{\sigma  - 1}}} \right]} \right)\,\ln \left[ {M\,\frac{{4\,\sigma  - 1}}{{\sigma  - 1}}} \right]} \right\}\,.
\end{align}
\normalsize
\indent Critical points of phase transitions can be identified by the singularities of the  Ruppeiner information curvature  (\ref{eqInfoCurvature2d}). The resulting expression is lengthy, but one can check that at $\sigma=-1/2$, which is the relevant divergence for the heat capacity (\ref{eqDSTHeatCapacity}), the Ruppeiner curvature is finite,  
\begin{equation}\label{eqDSTRuppeinerInfoCurvature}
\mathop {\lim }\limits_{\sigma  \to  - \frac{1}{2}} R_I^{(R)}(M,\sigma ) = \frac{{\left( {\ln \left( {{2^{9 + 7\,\ln 2}}\,{M^{3 + \ln (32\,M)}}} \right) - 4} \right)\,\ln (2\,M) + \ln {2^{{{\ln }^2}2 + 3\,\ln 2 - 4}} + 4}}{{8\,\pi \,M\,{{\ln }^2}\left[ {{2^{\ln 2}}{{(2\,M)}^{\ln (8\,M)}}} \right]}}\,,
\end{equation}
thus the Davies critical point cannot be covered by this particular thermodynamic geometry. The latter mismatch shows that the Ruppeiner information approach is not an appropriate choice for the description of the equilibrium state space of the DST black hole solution. 
\\
\indent Although Ruppeiner metric fails to produce a viable thermodynamic description, one can always impose only local thermodynamic stability defined by the positive values of the heat capacity $C>0$. The latter condition leads to the parameter region $-1<\sigma<-1/2$, together with $\sigma>1$ and arbitrary large mass $M>0$. 

\subsection{Weinhold information metric}

The Weinhold metric is defined as the Hessian of the mass of the black hole with respect to the entropy and the other extensive parameters. In the DST case one has
\begin{equation}
g_{ab}^{(W)} = {\partial _a}{\partial _b}M(S,\sigma )\,,\quad a,\,b = 1,2\,,
\end{equation}
where the mass $M$ of the DST black hole is given in terms of the entropy $S$ and the parameter $\sigma$ such as
\begin{equation}
M(S,\,\sigma ) = \frac{{\sigma  - 1}}{{2\,(4\,\sigma  - 1)}}\,{\pi ^{\frac{3}{{2\,(1 - \sigma )}} - 2}}\,{S^{\frac{3}{{2(\sigma  - 1)}} + 2}}\,.
\end{equation}
The heat capacity now looks quite simple,
\begin{equation}
C = \frac{{{\partial _S}M}}{{\partial _S^2M}}=\left( {1 - \frac{3}{{2\,\sigma  + 1}}} \right)\,S\,,
\end{equation}
and the Davies transition point at $\sigma=-1/2$ is present. The metric components are given explicitly by
\begin{align}
g_{SS}^{(W)} &= \frac{{{\pi ^{\frac{3}{{2 - 2\,\sigma }} - 2}}(2\,\sigma  + 1)\,{S^{\frac{3}{{2\,(\sigma  - 1)}}}}}}{{8\,(\sigma  - 1)}}\,,\\\nonumber
\\
g_{S\sigma }^{(W)} &= g_{\sigma S}^{(W)} =  - \frac{{3\,{\pi ^{\frac{3}{{2 - 2\,\sigma }} - 2}}\,{S^{\frac{3}{{2\,(\sigma  - 1)}} + 1}}\,\ln \left( {{\pi ^{1 - 4\sigma }}{S^{4\sigma  - 1}}} \right)}}{{8\,{{(\sigma  - 1)}^2}\,(4\,\sigma  - 1)}}\,,\\\nonumber
\\\nonumber
g_{\sigma \sigma }^{(W)} &= \frac{{3{\pi ^{\frac{3}{{2 - 2\sigma }} - 2}}{S^{\frac{3}{{2(\sigma  - 1)}} + 2}}}}{{8{{(\sigma  - 1)}^3}{{(4\sigma  - 1)}^3}}}\,\left[ {32 + 4\,(\sigma  - 1)\,\left( {8\,{{(\sigma  - 1)}^2} + 3\,(1 - 4\,\sigma )\,\ln \pi } \right)\,\ln 2} \right.\\\nonumber
 &+ 96\,{\sigma ^2} - 32\,{\sigma ^3} - 96\,\sigma  + 12\,{(\sigma  - 1)^2}\,{\ln ^2}2 + 3\,{(1 - 4\,\sigma )^2}\,{\ln ^2}\pi  - 16\,{(\sigma  - 1)^2}\,(4\,\sigma  - 1)\,\ln \pi \\\nonumber
& - 4\,(\sigma  - 1)\,\ln (\sigma  - 1)\,\left( {2\,(\sigma  - 1)\,(4\,\sigma  - 4 + \ln 8) + \,\ln \left( {{\pi ^{3\,(1 - 4\,\sigma )}}\,{{(\sigma  - 1)}^{3\,(1 - \sigma )}}} \right)} \right)\\\nonumber
& + 2\,(4\,\sigma  - 1)\,\left( {2\,\sigma \,\left( {4\,\sigma  - 8 + \ln \left( {\frac{8}{{\,{\pi ^6}}}} \right)} \right) - \ln \left[ {\frac{{{\pi ^3}}}{{64}}\,{{(\sigma  - 1)}^{6\,(\sigma  - 1)}}} \right] + 8} \right)\\
&\left. {\times \ln \left( {{{\left( {\frac{{\sigma  - 1}}{2}} \right)}^{\frac{{2\,(\sigma  - 1)}}{{4\,\sigma  - 1}}}}\,S} \right) + 3\,{{(1 - 4\,\sigma )}^2}\,{{\ln }^2}\left( {{{\left( {\frac{{\sigma  - 1}}{2}} \right)}^{\frac{{2\,(\sigma  - 1)}}{{4\,\sigma  - 1}}}}\,S} \right)} \right]\,.
\end{align}

The Weinhold approach also fails to reproduce the Davies transition point due to the fact that the Weinhold curvature is finite at $\sigma=-1/2$,
\begin{equation}
\mathop {\lim }\limits_{\sigma  \to  - 1/2} {R^{(W)}_I}(M,\,\sigma) =  - \frac{{6.28319\,(\ln S - 3.14473)\,\left( {{{\ln }^2}S - 3.28946\,\ln S + 4.45514} \right)}}{{S\,{{(\ln S - 1.14473)}^4}}}\,,
\end{equation}
taking into account that $S>0$ is everywhere assumed.

\section{Legendre invariant thermodynamic geometries on the equilibrium state space of the DST black hole solution }\label{sec:6}

\subsection{Quevedo information metric}
\begin{figure}[H]
    \centering
    \begin{subfigure}[b]{0.45\textwidth}
        \includegraphics[width=\textwidth]{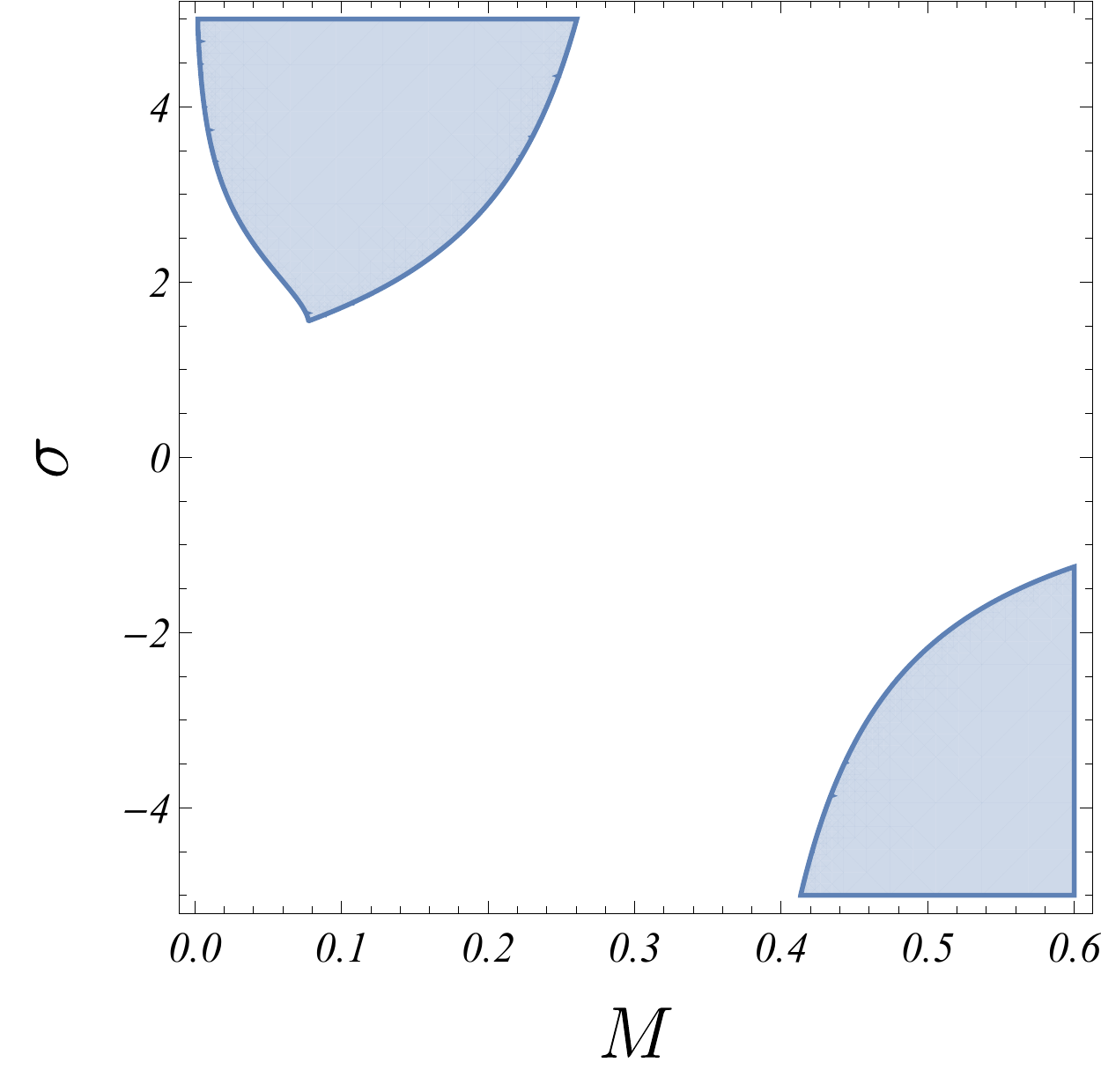}
    \end{subfigure}
    \caption{The regions of positive definite information metric together with $C>0$ (the shaded regions) for the DST black hole with respect to the Quevedo information approach. The upper region lies within $1< \sigma< \infty$ and $0<M<1/3$, while the lower region is defined within $-\infty<\sigma<-1/2$ and $1/3<M<\infty$.}
\label{figDSTmetricQ}
\end{figure}

The Quevedo information metric on the equilibrium state space of the DST solution is given by
\begin{equation}
ds_Q^2 = {\beta _S}\,S\,(\partial _\sigma ^2S\,d{\sigma ^2} - \partial _M^2S\,d{M^2}) = {{g}^{(Q)}_{MM}}\,d{M^2} + {{g}^{(Q)}_{\sigma \sigma }}\,d{\sigma ^2}\,.
\end{equation}
One can find the degree of generalized homogeneity, $\beta_S$, directly from Euler's theorem for homogeneous functions (\ref{eqEulerIdentity}):
\begin{equation}
M\,\frac{{\partial S}}{{\partial M}} + \sigma \,\frac{{\partial S}}{{\partial \sigma }} = {\beta _S}\,S\,.
\end{equation}
The latter equation for $\beta_S$ leads to the following components of the information metric:
\begin{equation}
{{g}^{(Q)}_{MM}} =  - (M\,{\partial _M}S + \sigma \,{\partial _\sigma }S)\,\partial _M^2S\,,\qquad {{g}^{(Q)}_{\sigma \sigma }} = (M\,{\partial _M}S + \sigma \,{\partial _\sigma }S)\,\partial _\sigma ^2S\,.
\end{equation}
The regions of positive definite information metric, together with local thermodynamic stability $C>0$, in Quevedo's case are shown on Figure \ref{figDSTmetricQ}. The upper region is constrained within $\sigma>3/2$ and $0<M<1/3$, while the lower region lies within $\sigma<-1/2$ and $M>1/3$. One should have in mind that contrary to the Ruppeiner's case, in Quevedo's case we do not have clear physical interpretation of the components of the information metric, thus one is not compelled to impose the Sylvester criterion. The latter will not necessarily give the regions of global thermodynamic stability. On the other hand, one can check that the convexity condition, $\partial_a\partial_b S \ge 0$, cannot be satisfied here. 
\indent The Quevedo thermodynamic curvature on the two-dimensional manifold ($M,\,\sigma$)\footnote{For clarity we have omitted the superscript (Q) from the metric components.},
\begin{align}
R_I^{(Q)} = \frac{{g_{MM}\left( {g_{MM,\sigma }^{}g_{\sigma \sigma ,\sigma }^{} + g_{\sigma \sigma ,M}^2} \right) + g_{\sigma \sigma }^{}\left( {g_{MM,\sigma }^2 + g_{MM,M}^{}g_{\sigma \sigma ,M}^{} - 2g_{MM}^{}\left( {g_{MM,\sigma ,\sigma }^{} + g_{\sigma \sigma ,M,M}^{}} \right)} \right)}}{{2\,g_{MM}^2\,g_{\sigma \sigma }^2}}\,,
\end{align}
is singular at $M= 1/3$, and it is also singular at the Davies transition point $\sigma\to -1/2$, suggesting that Quevedo information metric is an appropriate metric for the description of the equilibrium state space of the DST solution. However, one can check that there are also additional spinodal curves.
\subsection{HPEM information metric}
\begin{figure}[H]
    \centering
    \begin{subfigure}[b]{0.45\textwidth}
        \includegraphics[width=\textwidth]{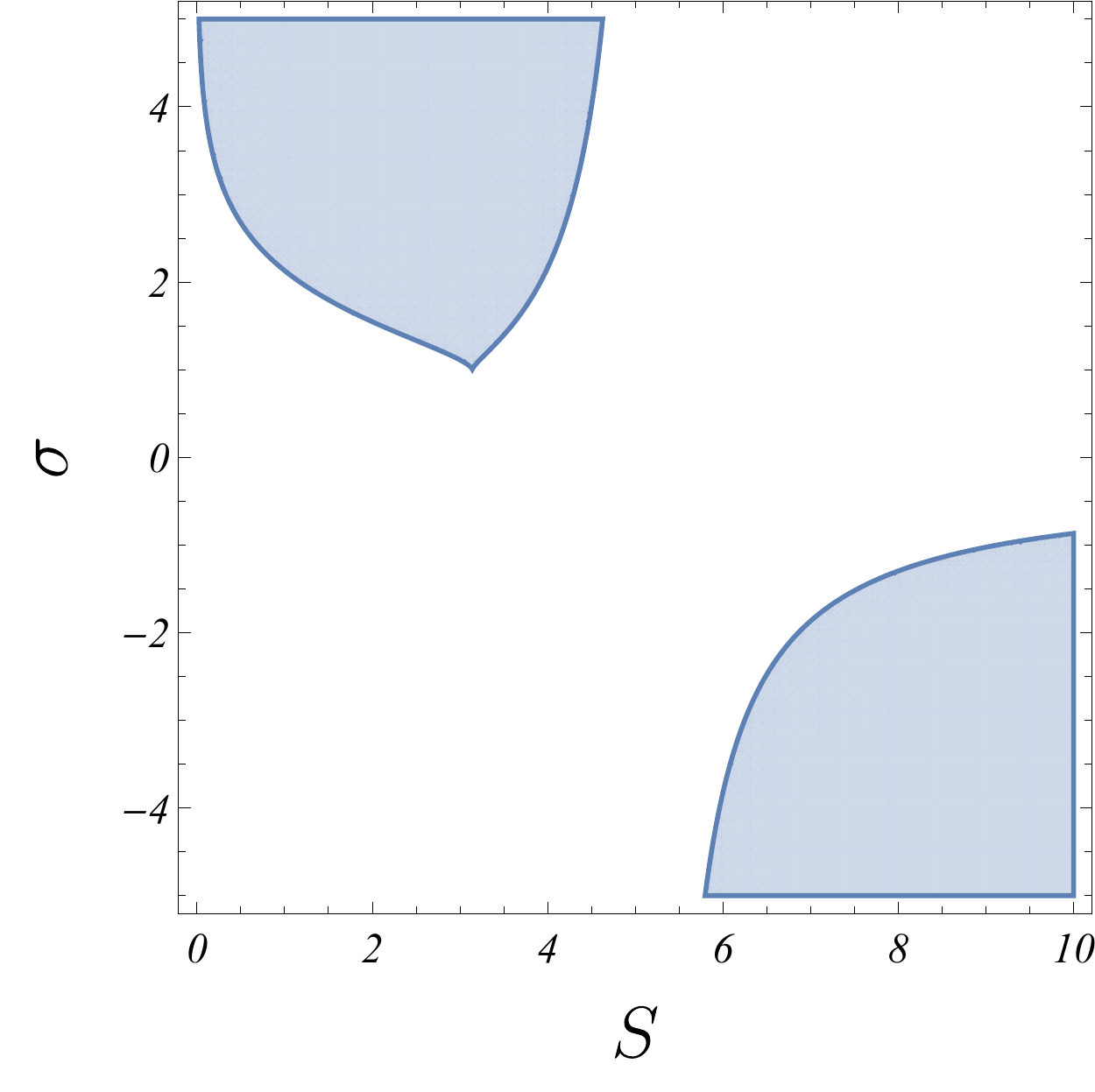}
    \end{subfigure}
    \caption{The regions of positive definite information metric together with $C>0$ (the shaded regions) for the DST black hole with respect to the HPEM information metric.  The upper region lies within $1< \sigma< \infty$, while the lower region is defined within $-\infty<\sigma<-1/2$.}
\label{figDSTmetricHPEM}
\end{figure}

In order to avoid extra singular points in the Quevedo thermodynamic curvature, which do not coincide with phase transitions of any type, in \cite{Hendi:2015rja} the authors proposed an alternative information metric with different conformal factor,
\begin{equation}
ds_{HPEM}^2 = S\,\frac{{{\partial _S}M}}{{{{(\partial _\sigma ^2M)}^3}}}\,( - \partial _S^2M\,d{S^2} + \partial _\sigma ^2M\,d{\sigma ^2})\,,
\end{equation}
with components
\begin{equation}
g_{SS}^{(HPEM)} =  - S\,\partial _S^2M\,\frac{{{\partial _S}M}}{{{{(\partial _\sigma ^2M)}^3}}}\,,\qquad g_{\sigma \sigma }^{(HPEM)} =  S\,\partial _\sigma ^2M\,\frac{{{\partial _S}M}}{{{{(\partial _\sigma ^2M)}^3}}}\,.
\end{equation}
One can find the regions where the Sylvester's criterion holds together with $C>0$ as shown on Fig. \ref{figDSTmetricHPEM}. 
\indent The HPEM information curvature\footnote{For clarity we have omitted the superscript (HPEM) from the metric components.},
\begin{align}
R_I^{(HPEM)}(S,\,\sigma ) = \frac{{g_{SS}^{}\left( {g_{SS,\sigma }^{}g_{\sigma \sigma ,\sigma }^{} + g_{\sigma \sigma ,S}^2} \right) + g_{\sigma \sigma }^{}\left( {g_{SS,\sigma }^2 + g_{SS,S}^{}g_{\sigma \sigma ,S}^{} - 2g_{SS}^{}\left( {g_{SS,\sigma ,\sigma }^{} + g_{\sigma \sigma ,S,S}^{}} \right)} \right)}}{{2\,g_{SS}^2\,g_{\sigma \sigma }^2}}\,,
\end{align}
is singular only at the Davies point $\sigma=-1/2$, thus HPEM metric is also an appropriate Riemannian metric on the equilibrium state space of the DST black hole. 

\subsection{MM information metric}
The final geometric approach, which we are going to consider, was proposed by A. H. Mansoori and B. Mirza in \cite{Mansoori:2013pna}. In the article the authors define a conjugate thermodynamic potential via an appropriate Legendre transformation. In the MM information approach the divergent
points of the specific heat turn out to correspond exactly to the singularities of the thermodynamic curvature. 
\\
\indent The conjugate potential we choose to work with is the Helmholtz free energy $F$, which is related to the mass $M$ by the following Legendre transformation
\begin{equation}
F(T,\,\sigma ) = M(T,\,\sigma ) - T\,S(T,\,\sigma )\,.
\end{equation}
The latter yields
\begin{equation}
F(T,\,\sigma ) =  - \frac{{2\,\sigma  + 1}}{{4\,\sigma  - 1}}\,{2^{2 - \frac{6}{{2\,\sigma  + 1}}}}\,{\pi ^{2 - \frac{3}{{2\,\sigma  + 1}}}}\,{T^{2 - \frac{3}{{2\,\sigma  + 1}}}}\,.
\end{equation}
The components of the MM thermodynamic metric are now given by
\begin{equation}
g_{TT}^{(MM)} = \frac{1}{T}\,\frac{{{\partial ^2}F}}{{\partial \,{T^2}}}\,,\qquad g_{T\sigma }^{(MM)} = g_{\sigma T}^{(MM)} = \frac{1}{T}\,\frac{{{\partial ^2}F}}{{\partial \,T\partial \sigma }}\,,\qquad g_{\sigma \sigma }^{(MM)} = \frac{1}{T}\,\frac{{{\partial ^2}F}}{{\partial \,{\sigma ^2}}}\,.
\end{equation}
One can show that there are no regions in the ($T,\,\sigma$) parameter space, where the Sylvester's criterion holds together with $C>0$. More importantly, the MM information curvature,
\begin{align}
\nonumber
&R_I^{(MM)}(T,\,\sigma ) = \frac{1}{{2\,{{\det }^2}( - \hat g)}}\\\nonumber
 &\times \left\{ {g_{TT}^{}\,\left[ {g_{TT,T}^{}\,g_{\sigma \sigma ,\sigma }^{} - 2\,g_{T\sigma ,\sigma }^{}\,\left( {g_{TT,\sigma }^{} - 2\,g_{T\sigma ,T}^{}} \right) - g_{\sigma \sigma ,T}^{}\,\left( {g_{TT,\sigma }^{} + 2\,g_{T\sigma ,T}^{}} \right)} \right]} \right.\\\nonumber
 &+ g_{TT}^{}\,\left( {g_{\sigma \sigma ,\sigma }^{}\,\left( {g_{TT,\sigma }^{} - 2\,g_{T\sigma ,T}^{}} \right) + g_{\sigma \sigma ,T}^2} \right) + 2\,g_{T\sigma }^2\,\left( {g_{TT,\sigma ,\sigma }^{} - 2\,g_{T\sigma ,T,\sigma }^{} + g_{\sigma \sigma ,T,T}^{}} \right)\\\nonumber
&\left. { + g_{\sigma \sigma }^{}\,\left[ {g_{TT,\sigma }^2 + g_{TT,T}^{}\,\left( {g_{\sigma \sigma ,T}^{} - 2\,g_{T\sigma ,\sigma }^{}} \right) - 2\,g_{TT}^{}\,\left( {g_{TT,\sigma ,\sigma }^{} - 2\,g_{T\sigma ,T,\sigma }^{} + g_{\sigma \sigma ,T,T}^{}} \right)} \right]} \right\}\,,
\end{align}
is singular exactly at the Davies transition point $\sigma\to -1/2$ without any extra critical points.

\section{Conclusion}\label{sec:conclusion}

Our current investigation is instigated by the intriguing existence of dark matter and dark energy in the Universe, which cannot be explained by traditional approaches. This motivates us to consider  alternative models, which can include effects related to these dark phenomena. Highly promising alternatives are the so called higher derivative theories of gravity, which include contributions from higher powers of the Ricci scalar or other geometric invariants. In particular, our focus is on the thermodynamic properties of their admissible black hole solutions, which will allow us to constrain the possible dark matter/energy contributions,  at least when thermodynamics is concerned. 
\\
\indent In this paper we consider one known four-dimensional higher derivative  black hole solution, namely the Deser-Sarioglu-Tekin black hole. The latter being a static, spherically symmetric gravitational solution of a theory with contributions from a non-polynomial term of the Weyl tensor to the Einstein–Hilbert Lagrangian. In order to study any implications for the black hole thermodynamics, we take advantage of two different geometric formulations, namely  those of the Hessian information metrics (geometric thermodynamics) and the formalism of Legendre invariant thermodynamic metrics (geometrothermodynamics) on the space of equilibrium states of the DST black hole. 
\\
\indent In general, the formalism of thermodynamic information geometry identifies the phase transition points of the system with
the singularities of the corresponding thermodynamic information curvature $R_I$. Near the critical points the underlying inter-particle interactions become strongly correlated and the equilibrium thermodynamic considerations are no longer applicable. In this case
one expects that a more general approach should hold.
\\
\indent In the Hessian formulation we analyzed the Ruppeiner and the Weinhold thermodynamic metrics and showed that they are inadequate for the description of the DST black hole equilibrium state space. This is due to the occurring mismatch between the singularities of the heat capacity and the singularities of the corresponding thermodynamic curvatures. Therefore the Hessian thermodynamic geometries are unable to reproduce the Davies type transition points of the DST black hole heat capacity. 
\\
\indent On the other hand, in the Legendre invariant case, all considered thermodynamic metrics successfully manage to incorporate the relevant phase transition points. Consequently they can be taken as viable metrics on the equilibrium state space of the DST black hole. However, some of them, such as the Quevedo metric, encounter additional singularities in their thermodynamic curvatures, the latter having obscure physical meaning at best. On the contrary, the HPEM and the MM metric seem to deal well with this problem and manage to get rid of the redundant spinodal curves in the case of the DST black hole.
\\
\indent Finally, let us address the problem of thermodynamic stability of the DST solution. For global stability one refers to Sylvester criterion for positive definite information metric, together with the positivity of the black hole heat capacity (local thermodynamic stability). Unfortunately, in the framework of geometric thermodynamics, both conditions can be interpreted as global thermodynamic stability only within the context of the Hessian metrics, due to their probabilistic interpretation. For the Legendre invariant metrics, imposing Sylvester criterion together with positive heat capacity does not necessarily guarantee global thermodynamic stability. The latter is caused by the current lack of physical interpretation of the components of the corresponding information metrics. Therefore, due to the failure of Hessian geometries, the DST black hole is only stable locally from a thermodynamic standpoint. The condition for local thermodynamic stability, together with the divergences of the physical DST metric curvature, constrain the values of the unknown parameter $\sigma$ in the regions $\sigma<-1/2$ and $\sigma>1$. The latter is also confirmed by imposing the Sylvester criterion for Quevedo and HPEM thermodynamic metrics.

\section*{Acknowledgements}

The author would like to thank R. C. Rashkov, H. Dimov, S. Yazadjiev and D. Arnaudov for insightful discussions and for careful reading of the draft. This work was supported by the Bulgarian NSF grant \textnumero~DM18/1 and Sofia University Research Fund under Grant \textnumero~80-10-104.







\bibliographystyle{utphys}
\bibliography{TS-refs}

\end{document}